\documentclass{iau-JDSS}
\usepackage{graphicx}

\newcommand{\hi}{{\rm H{\textsc{i}}\,}}
\newcommand{\hii}{{\rm H{\textsc{ii}}\,}}

\newcommand{\cii}{{\rm C{\textsc{ii}}\,}}

\newcommand{\civ}{{\rm C{\textsc{iv}}\,}}

\newcommand{\oi}{{\rm O{\textsc{i}}\,}}

\newcommand{\oiii}{{\rm O{\textsc{iii}}\,}}

\newcommand{\ovi}{{\rm O{\textsc{vi}}\,}}

\title[ISM Simulations: An Overview of Models]{ISM Simulations: An Overview of Models}

\author[Avillez et al.]{M.A. de Avillez$^{1,2}$, D. Breitschwerdt$^2$, A.
Asgekar$^{3}$, and E. Spitoni$^1$} 

\affiliation{$^1$Dept. of Mathematics, University of \'Evora, 7000 \'Evora, Portugal \break
email: mavillez,spitoni@galaxy.lca.uevora.pt\\
[\affilskip] $^2$Dept. of Astronomy \& Astrophysics, Technical University Berlin, D-10623
Berlin, Germany \break email: breitschwerdt@astro.physik.tu-berlin.de\\
[\affilskip] $^3$ASTRON, P.O. Box 2, 7990 AA Dwingeloo, The
Netherlands}

\pubyear{2012}
\volume{Volume 16}  
\pagerange{xx--xx}
\date{?? and in revised form ??}
\jname{Highlights of Astronomy, Volume 16}
\editors{Chu Y.-H.}
\begin{document}

\maketitle

\begin{abstract}
Until recently the dynamical evolution of the interstellar medium (ISM) was simulated using
collisional ionization equilibrium (CIE) conditions. However, the ISM is a dynamical system, in
which the plasma is naturally driven out of  equilibrium due to atomic and dynamic processes
operating on different timescales. A step forward in the field comprises a multi-fluid approach
taking into account the joint thermal and dynamical evolutions of the ISM gas.

\end{abstract}

\firstsection 
\section{Introduction}

The attempts to model the supernova-driven ISM can be traced to the seminal models
of Cox \& Smith (1974; CS74) and McKee \& Ostriker (1977, MO77). In the former supernovae (SNe)
maintain an interconnected tunnel network filled with X-ray emitting gas, while in MO77 the gas is
distributed into three phases in global pressure equilibrium. In both models the Galactic volume
(50\% in CS74 and 70-80\% in MO77) is filled with hot ($>10^{5}$ K) low-density gas. Further
ramifications include the break-out of the hot intercloud medium, cooling and condensing into
clouds (galactic fountain; Shapiro \& Field 1976) or escaping as a wind (e.g., chimney model; Norman
\& Ikeuchi 1989).

Although these early works capture some of the essential physics, more complex and sophisticated
models were devised by taking advantage of numerical simulations. These comprise
the evolution of a patch of the Galactic disk in two dimensions (2D) (hydrodynamical (HD): Bania \&
Lyon 1980; Chiang \& Prendergast 1985; Chiang \& Bregman 1988; Rosen et al. 1993;
Magnetohydrodynamical (MHD): Vazquez-Semadeni et al. 1995), and in three-dimensions (3D), e.g.. the
MHD evolution of a $200^3$ pc$^{3}$ region (Balsara et al. 2004) and the cosmic-rays
driven amplification of the field in a differentially rotated domain ($0.5\times1\times [-0.6,0.6]$
kpc$^{3}$; Hanasz et al. 2004). The first disk-halo evolution models (2D~HD) were developed by Rosen
\& Bregman (1995). With increasing of computer power, 3D HD (de Avillez model in 2000 and upgrades -
see Avillez \& Breitschwerdt 2007; Joung \& Mac Low 2006) and MHD (Korpi et al. 1999; Avillez \&
Breitschwerdt 2005; Gressel et al. 2008; Hill et al. 2012) models have been developed. 

In general the disk-halo models consider parameters according to observations (e.g., initial matter
distribution with height, SN rates, background UV radiation field). Differences are
found in the number of physical processes included (magnetic fields, cosmic rays, heat
conduction, etc.), numerical techniques, type of grid (fixed or differentially rotated using the
shear box technique), and grid resolutions and sizes. Resolutions are fixed or benefit from
the use of the adaptive mesh refinement (AMR) technique (Berger \& Oliger 1984). The highest resolutions
cover a wide range from 0.5 pc to 10 pc, passing through 2 and 8 pc. The grid sizes in the vertical
direction range from 0.1 kpc to 15 kpc on either side of the Galactic midplane.
However, grids extending up to 2 kpc imply that the disk-halo-disk cycle can neither be
established nor tracked - the simulations are valid for a small period of time before
the gas escapes from the top and bottom boundaries. 

These simulations showed that: (i) the ISM does not become saturated by SN activity, (ii) the disk
expands and relaxes dynamically as SN rate fluctuates in time and space, (iii) the turbulent
field builds up exponentially within 20 Myr of disk evolution, (iv) the magnetic field does not
strongly correlate with density, except for the densest regions, (v) the magnetic field does not
prevent the matter escape into the halo as it only briefly delays the disk-halo cycle, (vi) the
volume filling factor of hot gas in the Galactic disk is only $\sim20$ \%, (vii) there are large
pressure variations in the disk in contrast to MO77 with the thermal pressure dominating at
high temperatures ($T>10^{6}$ K), magnetic pressure at $T<200$ K, and ram pressure elsewhere.

\section{Thermal \& Dynamical Evolution of the ISM}

All models referred previously assumed the ISM plasma to be in CIE, represented by a unique and
general cooling function (CF) taken from different sources (e.g., Dalgarno \& McCray 1972 (DM72);
Sutherland \& Dopita 1993; Gnat \& Sternberg 2007). CIE assumes that the number of ionizations is
balanced by recombinations from higher ionization stages. However, CIE is only valid provided the
cooling timescale ($\tau_{cool}$) of the plasma is larger than the recombination times scales of the
different ions ($\tau^{Z,z}_{rec}$), something that occurs at $T>10^{6}$ K (see references above).
For lower temperatures $\tau_{cool} <\tau_{rec}^{Z,z}$, and deviations from CIE are expected
(see, e.g., DM72). These departures affect the local
cooling, which is a time-dependent process that controls the flow dynamics, feeding back to the
thermal evolution by a change in the density and internal energy distribution, which in turn
modifies the thermodynamic path of non-equilibrium cooling.  

A major improvement in ISM studies is therefore to carry out time-dependent multi-fluid
calculations of the joint thermal and dynamical evolution of the plasma, i.e. to follow each fluid
element's thermal history by determining its ionization structure and CF at each time-step.
Radiative losses are folded into the energy equation with the internal energy including also
the potential energies associated to the different ionization stages. 

Historically, there have been a number of simulations, which have included part of the ionization
history into HD simulations, (Cox \& Anderson 1982; Innes et al. 1987; Borkowski et al. 1994;
Smith \& Cox 2001; among others), misty tailored for specific astrophysical problems. The effect of
delayed recombination has been emphasized by Breitschwerdt \& Schmutzler (1994), who have modelled
the soft X-ray background. Melioli et al. (2009), following the formation and evolution of \hi
clouds, only considered the time evolution of selected ions (\hi, \hii, \cii-\civ, and \oi-\oiii)
for temperatures below $10^{6}$ K, using a fit to the Sutherland \& Dopita (1993) CF for
$T\geq 10^{6}$ K. This setup has severe implications in the cooling of the gas as their calculation
does not trace the relevant ions recombining to \civ and \oiii.

Recently, owing to the development of the Atomic+Molecular Plasma Emission Code (EA+MPEC) and its
coupling to a PPM based AMR code, it has been possible to carry out multi-fluid calculations of the
ISM tracing both the thermal and dynamical evolutions of the gas self-consistently. The ionization
structure, cooling and emission spectra of H, He, C, N, O, Ne, Mg, Si, S, and Fe ions (with solar
abundances; Asplund et al. 2009) are traced on the spot at each time step assuming an equal
Maxwellian temperature for electrons and ions (see details and references in Avillez \&
Breitschwerdt 2012).

These simulations showed several interesting effects: (i) in a dynamic ISM, the ionization structure
and, therefore, the CF, varies with space and time, depending on the initial conditions and its
history, (ii) the cooling paths in general do not follow the one predicted by static plasma emission
calculations, (iii) non-equilibrium ionization X-ray emission in the $\sim 0.25$ keV band of gas
with $T<10^5$K can dominate the corresponding CIE emission at even $T=10^{6.2}$ K as a result of
delayed recombination, (iv) the presence of \ovi ions at temperatures $< 10^{5}$ K corresponding to
70\% of the total \ovi mass, and (v) a large fraction of electrons are found in the thermally
unstable regime and have a log normal distribution with similar properties (mean and dispersion) to
those derived from observations against pulsars with known distances.

\section{Conclusions}

The dynamical and thermal evolution of the ISM are strongly coupled, because the ionization
structure determines the CF, which in turn controls the dynamics and thereby the ionization
structure, closing a feedback loop. Consequently, strong deviations from CIE occur due to severe
mismatches between the different ionization/recombination and dynamical time scales of the
plasma. Similar effects due time-dependent cooling are expected in other astrophysical
contexts.

\begin{acknowledgments}
M.A.A. thanks the IAU and specifically Y.-H. Chu and V. Reuter for financial support.
\end{acknowledgments}

\end{document}